\documentclass[11pt]{article}
\usepackage{graphicx} 
\usepackage{amssymb}
\usepackage[colorlinks=true, linkcolor=blue, citecolor=blue, urlcolor=blue]{hyperref}
\usepackage{url}
\usepackage{authblk}
\usepackage{multirow}
\usepackage[letterpaper,top=2cm,bottom=3cm,left=2cm,right=2cm,marginparwidth=1.75cm]{geometry}
\usepackage[backend=bibtex, style=numeric-comp, sorting=none]{biblatex}
\newcommand{\pt}{\ensuremath{p_\mathrm{T}}}

\providecommand{\keywords}[1]
{
  \small	
  \textit{{Keywords---}} #1
}

\title{Jet observables in heavy ion collisions : a white paper}
\author[a]{Ankita Budhraja}
\affil[a]{\footnotesize Nikhef, Theory Group, Science Park 105, 1098 XG, Amsterdam, The Netherlands}
\author[a,b]{Marco van Leeuwen}
\affil[b]{Utrecht University, P.O. Box 80000, 3508 TA Utrecht, The Netherlands}
\author[c,d]{José Guilherme Milhano}
\affil[c]{Laborat\'orio de Instrumenta\c{c}\~ao e F\'isica Experimental de Part\'culas (LIP), \protect \\ Av.
    Professor Gama Pinto 2, 1649-003 Lisboa, Portugal}
\affil[d]{Departamento de F\'{\i}sica, Instituto Superior T\'ecnico,\protect \\
    Universidade de Lisboa, Av. Rovisco Pais 1, 1049-001 Lisboa, Portugal}
\date{\today}

\begin{document}

\maketitle

\begin{abstract}
    This paper presents an overview of a survey of jet substructure observables used to study modifications of jets induced by interaction with a Quark Gluon Plasma. We further outline ideas that were presented and discussed at the \textit{New jet quenching tools to explore equilibrium and non-equilibrium dynamics in heavy-ion collisions} workshop, which was held in February 2024 at the ECT$^{*}$ in Trento, Italy. The goal of this white paper is to provide a brief report on the study of jet quenching observables earlier conducted and to present new ideas that could be relevant for future explorations.
\end{abstract}


\section{Introduction and Motivation}

The heavy-ion programs at the Relativistic Heavy Ion Collider (RHIC) and at the Large Hadron Collider (LHC) have revealed properties of the strongly interacting matter created under extreme conditions~\cite{Gyulassy:2004zy,Muller:2006ee,Muller:2012zq}. This strongly interacting matter, commonly referred to as the quark-gluon plasma (QGP), provides a rich opportunity to study deconfined quarks and gluons. Additionally, this new state of matter exposes a unique opportunity to study the phase diagram of Quantum Chromodynamics (QCD). However, since this state of matter exists for only about ${\cal O}(10^{-23}) \mathrm{s}$ after the collision, one needs to look for natural probes that originate during these collisions in order to study it.  
One such approach is to examine the evolution of highly energetic quarks and gluons, generated at the hard interaction of the two nuclei, as they traverse the QGP medium. These energetic partons shower into collimated sprays of particles referred to as jets. By studying the modifications of jets in the QGP, one aims to uncover the microscopic interactions of quarks and gluons with the medium and thereby the properties of the medium itself. An important experimental evidence that has been observed is the suppression of jets in the medium (even for very high $p_T$ jets) when compared to the vacuum baseline (p-p collisions)~\cite{ATLAS:2014ipv,CMS:2016uxf,ATLAS:2018gwx,ALICE:2019qyj}. 
This serves as a prominent evidence of the matter that is created. A wide range of efforts has been 
directed at formulating an understanding of the modification in the structure of these jets as they traverse through the QGP~\cite{Salgado:2003rv,Zapp:2008gi,Casalderrey-Solana:2014bpa,He:2015pra,Chien:2015hda,Cao:2017hhk,Caucal:2019uvr,Casalderrey-Solana:2019ubu,Vaidya:2020cyi,Vaidya:2020lih,Cunqueiro:2021wls,Caucal:2021cfb,Mehtar-Tani:2021fud,Cao:2022odi,Budhraja:2023rgo,Barata:2023bhh,Zhang:2023oid,Andres:2023xwr,Andres:2024ksi,Singh:2024vwb,Bossi:2024qho}.  

One of the most exciting advances in the study of jets, both theoretically and experimentally, has been the development of techniques to study the internal structure of jets to determine the properties of the underlying microscopic collisions.
Jet substructure studies in heavy-ion collisions aim at disentangling the properties of the QGP by looking at the modifications of the jets' inner structure in A-A collisions when compared to p-p collisions. However, due to the extraordinary complexity of the system created in these collisions as well as the presence of very large backgrounds, the simple task of relating the modifications in the substructure of jets to medium properties is far from trivial. 
In addition, while medium-induced gluon radiation is the main topic of physical interest, additional effects like colour coherence, elastic scattering and medium response, are expected to contribute to the observed modifications of jets and their structure. To facilitate an understanding of the medium properties, observables to study jet quenching in heavy-ion collisions should be selected considering two main aspects: sensitivity to the specific physics aspect of interest and the robustness of the observable against underlying event background. Some examples of the specific physics aspects that a given observable may be sensitive to are listed as under:

\begin{itemize}
\item \textit{The angular distribution of radiation pattern inside a jet}: The multiple soft kicks received by partons in the developing shower lead to a broadening of the radiation (resolved by the medium) inside the jet in addition to the presence of additional medium-induced radiation, see Section \ref{sec:structure} below;

\item \textit{Azimuthal broadening of partons as well as rare hard momentum exchanges with the medium (Moli\`ere scattering)}: In fact, the angular broadening can be directly related to the jet transport coefficient $\hat{q}$, see Section \ref{sec:structure} below;

\item \textit{Differences between quark vs gluon-initiated  jets}: Naively, one expects gluons to interact more strongly with the medium and hence gluon-initiated jets to lose more energy than quark-initiated jets. Designing observables sensitive to such differences could, therefore, help provide a more differential understanding towards the energy loss mechanisms in the medium, see Ref.~\cite{Pablos:2022mrx} for one such proposed jet observable. Furthermore, observables that can be made sensitive to heavy quark mass effects such as the dead cone can be of particular interest as well.
  
\item \textit{Path length dependence of parton energy loss}: Partons travelling along directions in which the medium is larger will, in principle, interact more with the medium and lose energy differently than partons travelling in directions where the medium extent is shorter.  

\item \textit{Interference effects in the medium}: If the formation time of a radiation is smaller than the distance between the scatterers (incoherent emissions), the radiation is resolved by the medium and can be considered as an independent source for further emissions while if the formation time is larger than the distance between the scattering centers (coherent emissions), multiple scatterers act coherently until the radiation is resolved by the medium. 

\item \textit{Medium response}: the back reaction from the medium to jet propagation. 

\item \textit{Formation time of emissions} that can be used to tag splittings that happened inside the medium from those outside.
  
\end{itemize}

The other important consideration is the robustness of the measured value of a jet observable against contributions from the underlying event background. In heavy-ion collisions, a large number of particles are produced, most of which are not associated to a hard scattering. In data analysis, the $p_\mathrm{T}$-density of the background is estimated on an event-by-event basis using $\eta-\phi$ areas outside the jet cone and  subtracted. However, statistical fluctuations of the background level inside the jet cone are significant. Some background subtraction methods attempt to provide an event-by-event estimate of the in-cone background \cite{Haake:2018hqn,ALICE:2023waz,Mengel:2024fcl}. 
Furthermore, as jets in vacuum are highly collimated, 
this makes it particularly interesting to study medium-induced radiation and medium response~\cite{Cao:2020wlm} appearing typically at large angles, where these effects may dominate over the vacuum parton shower physics. Unfortunately, these effects compete with the background effects in particular at large angles and an excellent control of the background subtraction is needed to study such large angle radiation. 

This article is organized as follows. In Section~\ref{sec:structure}, we present some general aspects of physics of jet modification in the medium. In Section~\ref{sec:survey}, we outline the major findings of the survey of jet observables conducted in Ref.~\cite{CrispimRomao:2023ssj}. This analysis aims towards finding the minimal set of observables that provides mutually uncorrelated information about the medium. In Section~\ref{sec:gamma-jet} and \ref{sec:future}, we discuss observables for future measurements that were discussed at the ECT$^{*}$ meeting.\footnote{The ECT$^{*}$ meeting page can be found at this link: \url{https://indico.ectstar.eu/event/198/timetable}.} Finally, in Section~\ref{sec:background}, we discuss the resilience of jet substructure observables to background contributions in heavy-ion collisions.

\section{Angular and longitudinal structure}
\label{sec:structure}

Hard scattering processes in the initial state of collisions of hadrons and nuclei produce quarks and gluons with a large transverse momentum. These highly energetic partons radiate as they propagate outward, forming parton showers, which subsequently hadronize, giving rise to jets of high-momentum particles in the final state. Jets are identified in experiments using jet finding algorithms which are formulated in a manner that they are insensitive to soft and collinear structures of the radiation, typically associated with divergences in the theoretical treatment.  As a result, the total energy and momentum of a jet is a measure of the energy and momentum of the parton (quark or gluon) that initiates it. In heavy-ion collisions, interactions of energetic partons with the thermal medium induce additional radiation due to elastic as well as inelastic interactions of the parton shower with the medium. A specific feature of the medium-induced radiation is the absence of any collinear enhancement associated to it. As a result of this, medium-induced radiation is typically at large angles. Gluons (or quarks) emitted at such large angles may escape the jet cone, leading to \textit{energy loss}. In addition, the internal structure of the jet is also modified by the jet-medium interactions. 

Generically speaking, the expected effects of jet quenching on the jet structure are: (1) a softening of the longitudinal structure and (2) 
a broadening of its transverse structure. The softening of the longitudinal structure is directly driven by additional medium-induced splittings which reduce the number of fragments with a large momentum fraction $z$ and increase the number of fragments at low $z$. Medium response, i.e. soft partons from the QGP that acquire momentum due to interactions with the jet and end up in the jet cone, may also contribute to an increase of the number of soft fragments. The transverse broadening of the jet structure, on the other hand, is driven by two separate effects: medium-induced radiation (and the corresponding recoil of the jet axis), as well as by momentum broadening due to transverse momentum exchanges with the medium partons (elastic scattering). It is important to note that these effects are with respect to reference jets in p-p collisions produced by partons with the same (initial) energy that fragment without interacting with the medium. In experiments, jet properties are generally compared between p-p and A-A collisions at the same reconstructed jet energy. In A-A collisions, a jet in a given observed p$_T$ range may originate from a parton with higher momentum that has lost a significant amount of energy due to out-of-cone radiation or a parton with only slightly higher momentum that lost a small amount of energy. This leads to a selection bias effect for jets in heavy-ion collisions. The resulting biases can, in principle, be reduced or avoided by either using $\gamma$-jet and Z-jet pairs (see Section~\ref{sec:gamma-jet} below), or through the procedure outlined in \cite{Brewer:2018dfs}.

The understanding of jet quenching effects in the medium is then largely driven by the study of modifications to the internal structure of jets. Jet substructure can be characterised by reporting distributions of the longitudinal and transverse (opening angle) distributions, or their moments, called {\it angularities}. It is worth noting that while fragment distributions provide an inclusive measure of the jet structure, jet shape variables like the angularities provide a measure for each jet, which may be sensitive to event-by-event variations in path length or energy loss. Another strategy for characterising the jet structure involves going back in the clustering history of the jet and extracting characteristic variables; this is often combined with grooming techniques that aim to provide a robust measure of the 'hardest' jet substructure, e.g. by reporting the momentum fraction and/or opening angle of a hard splitting ($z_g$, $R_g$) \cite{Mehtar-Tani:2019rrk, Caucal:2019uvr}. Additionally, dynamical grooming with different orderings can be utilized to obtain variables like the $k_{\perp}$ of the hardest splitting or even to access time structure of the splittings in heavy ion collisions~\cite{Apolinario:2020uvt,Pablos:2022mrx}. 

\section{Survey of observables: main results}
\label{sec:survey}

Heavy-ion collisions are a complex experimental environment, which also means that individual measurements are generally sensitive to a combination of different physics effects. To identify the jet observables that are most sensitive to jet quenching effects in the medium, a survey of 31 jet observables was conducted, using the JEWEL model as a reference for quenching effects.
The full results are presented in \cite{CrispimRomao:2023ssj} and we provide a short summary of the main findings here for reference.

The observables considered are listed in Table~\ref{tab:vars}. All the jet observables are calculated based on the constituents of a groomed jet.
\begin{table}[h]
    \centering
    \begin{tabular}{l|l}
        \hline\hline
        Observable                                                              & Type                                                       \\ \hline
        $y_{SD}$                                                                & \multirow{5}{*}{Jet Momenta and  Constituent Multiplicity} \\
        $\phi_{SD}$                                                             &                                                            \\
        $\Delta p_{T,SD} = p_{T,jet} - p_{T,jet_{SD}}$                          &                                                            \\
        $m_{SD}$                                                                &                                                            \\
        $n_{{\rm const},SD}$                                                             &                                                            \\ \hline
        $\bar r_{SD} = \frac{1}{n_{{\rm const},SD}}\lambda^0_{1,SD}$                     & \multirow{6}{*}{Angularities}                              \\
        $\bar r^2_{SD} = \frac{1}{n_{{\rm const},SD}}\lambda^0_{2,SD}$                   &                                                            \\
        $rz_{SD} = \lambda^1_{1,SD}$                                            &                                                            \\
        $r^2z_{SD} = \lambda^1_{2,SD}$                                          &                                                            \\
        $\bar z^2_{SD} = \frac{1}{n_{{\rm const},SD}} \lambda^2_{0,SD}$                  &                                                            \\
        $p_TD_{SD} = \sqrt{\sum_{i\in jet_{SD}} p^2_{T,i}} / p_{T,jet,SD}$      &                                                            \\ \hline
        $\tau_{2,SD},\ \tau_{3,SD}$                                             & \multirow{2}{*}{$N$-subjettiness}                          \\
        $\tau_{1,2,SD},\ \tau_{2,3,SD}$                                         &                                                            \\ \hline
        $|Q^{0.3}_{SD}|,\  |Q^{0.5}_{SD}|,\  |Q^{0.7}_{SD}|, \ |Q^{1.0}_{SD}|$, & Jet-Charges                                                \\ \hline
        $R_g,\ z_g, \ n_{SD}$                                                   & SoftDrop Grooming Intrinsic                                \\ \hline
        $R_{g, A}, \ z_{g, A}, \ \kappa_{A}$ with $A\in\{TD, ktD, zD\}$         & Dynamical Grooming Intrinsic                               \\
        \hline\hline
    \end{tabular}
    \caption{\label{tab:vars}Overview of the set of observables considered in the analyses computed on the constituents of a softdrop ($SD$) groomed jet.~\cite{CrispimRomao:2023ssj}}
\end{table}
Two complementary machine learning-based analyses are employed to study the correlations between these observables separately for p-p and A-A jet samples: linear correlations are studied using Principal Component Analysis (PCA) and non-linear relations are studied using a Deep Auto-Encoder (AE). Below we list the main findings of this survey :
\begin{enumerate}

\item The information content of the entire set is described by a small number of effective degrees of freedom. For the PCA, this effective degrees of freedom corresponds to the first 10 principal components describing about $\sim$ 90\% of the distributions of all input observables. When the non-linear relations are also included through the AE analysis, this further reduces to only 5 nodes on the main hidden layers providing similar reconstruction ability.
It is important to note that these effective degrees of freedom do not necessarily correspond to simple observables. Instead these may correspond to a subset or a combination of all the input observables with suitable weights.

\begin{figure}
    \centering
    \includegraphics[width=.5\textwidth]{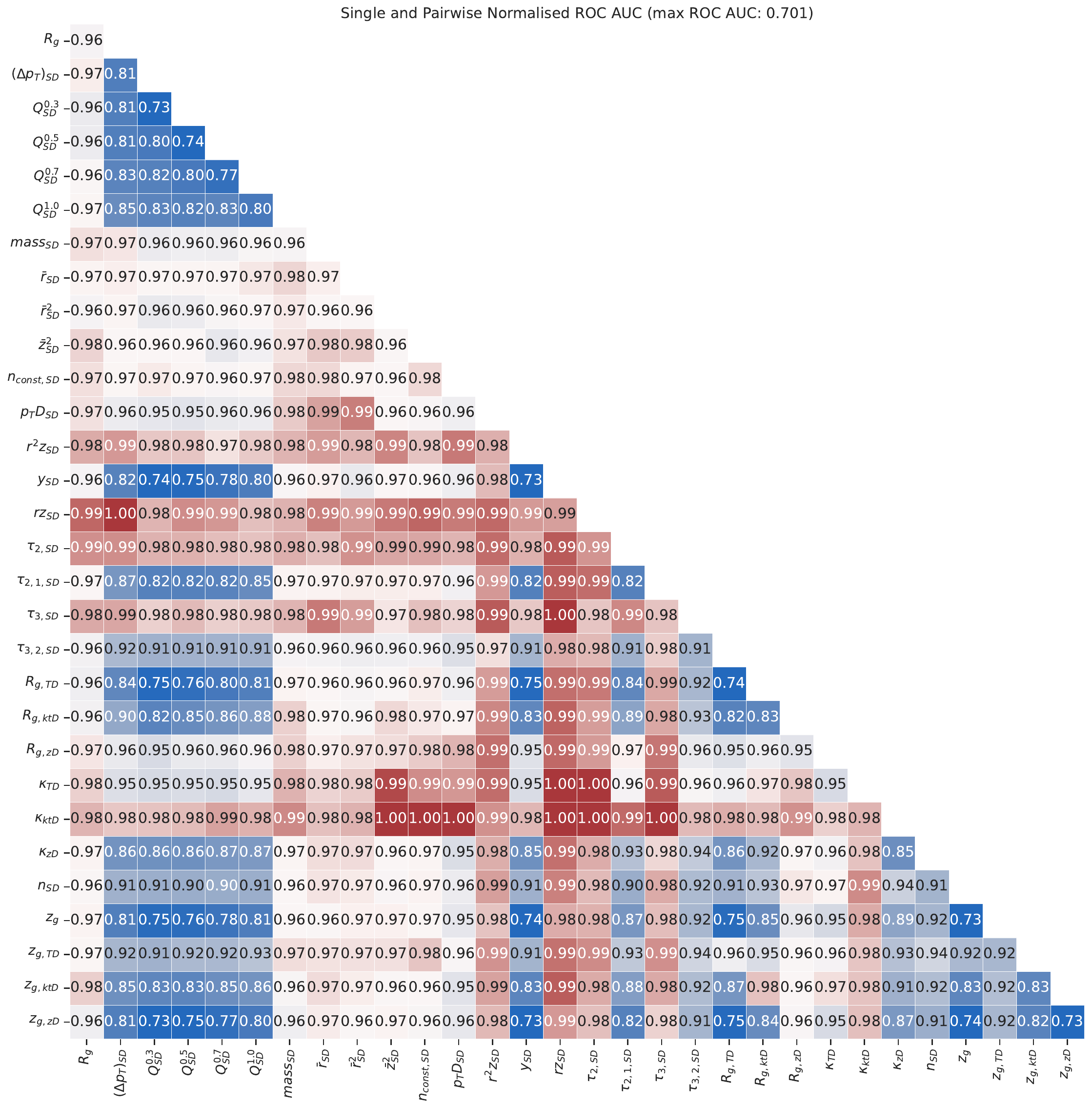}
    \caption{\label{fig:pairwise_rocs} Single and pairwise ROC area under curve (AUC) normalised to the ROC AUC obtained using all observables. Values close to $1$ (dark red boxes) signify identical discrimination ability as the full BDT trained on all observables. From \cite{CrispimRomao:2023ssj}.}
\end{figure}

\item The correlations between observables are mostly resilient to quenching effects included in the JEWEL model. This is found for both the linear and non-linear analyses: in both cases, the ability to reconstruct the observables in A-A samples using the effective degrees of the p-p sample indicates remarkable resilience of  correlations to quenching effects. The effect of quenching is manifested through a strong population migration modifying mostly the mean or most probable values of observables and not the correlation between pairs. 

\item Specific observables and pairs of observables provide similar discrimination ability as the full set. By training a boosted decision tree (BDT) on all observables and comparing it to the discrimination ability achieved by BDTs trained on each single observable and pairs of these observables, it is found that several observables as well as pairs of observables already exhaust the discrimination ability of the full BDT.

\end{enumerate}

More specifically, from Figure~\ref{fig:pairwise_rocs}, we find that the individual observables that are the most senstive to jet quenching in this study are $rz_{SD}$ and $\tau_{2, SD}$, each accounting for 0.99 of the discriminating power of the BDT trained in all observables. Similarly, $n_\mathrm{const,SD}$, $r^2z_{SD}$, $\tau_{3, SD}$, $\kappa_{ktD}$ are equally powerful observables, reaching 0.98 of the discriminating power. Additionally, some pairs of observables match the discrimination power of the BDT trained in all observables. These are pairings of $rz_{SD}$ with $(\Delta p_T)_{SD}$, $\tau_{3, SD}$, $\kappa_{TD}$ or $\kappa_{ktD}$; also the further pairings of $\kappa_{ktD}$ with any of $n_\mathrm{const,SD}$, $p_TD_{SD}$, $\bar z^2_{SD}$, $\tau_{2, SD}$ or $\tau_{3, SD}$: and $\kappa_{TD}$ with $\tau_{2, SD}$. Pairs involving a dynamical grooming observable and an angularity-type observable dominate this list.

\section{Reducing the biases: $\gamma$-jet measurements}
\label{sec:gamma-jet}
The energy loss of a parton propagating through a given amount of QGP has very large fluctuations. In most experimentally relevant situations, the number of medium-induced radiations is small and Poisson fluctuations of this number already lead to a large spread in energy loss, with a sizeable probability of no energy loss \cite{Salgado:2003gb,Armesto:2011ht}. Moreover, medium-induced radiation may look similar to vacuum radiation, and/or be partially recaptured by the jet finding algorithm. 
As discussed briefly in Section~\ref{sec:structure}, when selecting a given momentum range for data analysis or theoretical consideration from an inclusive sample of jets, the sample includes a combination of jets that did not lose energy in the QGP and jets that did lose energy and originate from partons with a larger initial transverse momentum. Due to the steeply falling jet spectrum, the contribution of partons with large energy loss is naturally suppressed; this effect is sometimes referred to as the \textit{leading jet bias} or \textit{fragmentation bias}, to indicate that an inclusive sample is generally biased towards jets that did not lose much energy and therefore fragment like jets in p-p collisions. This bias can be reduced by using experimental techniques that use jet pairs or gauge boson-jet pairs to gain access to the initial parton energy.

The clearest access to the initial parton kinematics is provided by the use of $\gamma$-jet and $Z$-boson-jet pairs. At leading order, a perfect balance of the boson and jet momenta is expected: the transverse momentum of the gauge boson is equal to that of the jet. The transverse momentum of photons and $Z$-bosons can be measured in experiment and since they do not lose energy in the QGP, this provides a direct measure of the initial jet parton momentum. Measurements so far have focused on $\gamma$-jet momentum balance \cite{ATLAS:2018dgb}, as well as measurements of both longitudinal and transverse fragment distributions in the recoil jet~\cite{CMS:2018mqn, CMS:2018jco}. These momentum balance measurements show a clear increase of the number of asymmetric pairs, where the reconstructed jet momentum is significantly smaller than the photon momentum. In the currently available analyses, the combination of the photon momentum selection and the minimum jet \pt{} result in a relatively large cut-off in the energy asymmetry, e.g. $x_{j/\gamma}=p_\mathrm{T,jet}/p_\mathrm{T,gamma} > 0.3$ for $p_T^{\gamma} = 100 - 158\, {\rm GeV}$ see Ref.~\cite{ATLAS:2018dgb}, leading to a significant loss of pairs. With larger data samples in Run 3 and 4 of the LHC, the photon energy range can be increased and the asymmetry selection can be relaxed to reduce the bias on the recoil jet sample. This was discussed in two separate presentations at the workshop~\cite{ECT_cunqueiro, ECT_lee}.

\section{Novel observables for future studies}
\label{sec:future}
 
With the goal to better understand the physics of jet quenching mechanisms, several suggestions were made during the meeting to propose observables/measurements that can reduce the biases in jet selection. These observables broadly fall into the following classes: energy flow-based, multivariate observables, and $\gamma$/Z-tagged observables (see Section~\ref {sec:gamma-jet}).  
Proposals were also made to study asymmetry-based observables that would, in principle, be more robust against background effects than the standard jet observables providing a complementary handle towards properties of the medium. Below we will outline these ideas in more detail. 

The first class of observables concerns energy-energy correlators (EECs) that describe asymptotic energy flow in scattering events and exhibit fundamentally distinct properties from standard jet substructure observables, like the ones considered in Table~\ref{tab:vars}. Correlation functions are one of the most fundamental objects in a field theory and enjoy a direct description in terms of the light ray operators. This enables a direct theory-experiment comparison to be possible. On the theoretical front, due to the simple analytic structure of EECs, their vacuum baseline is extremely well controlled with the 2-point energy correlator known precisely up to N$^3$LL accuracy and up to NLL for the higher-point correlators~\cite{Dixon:2019uzg, Gao:2023ivm, Chen:2020adz}. Furthermore, the energy weighing in the observable naturally suppresses the sensitivity to soft radiation without the need for jet grooming techniques. The presence of an accurate baseline as well as reduced soft sensitivity of the observable make it an interesting candidate for jet quenching studies in the medium. A few recent efforts have been dedicated to calculating the 2-point EECs in the medium~\cite{Andres:2022, Andres:2023xwr, Andres:2024ksi,Yang:2023dwc,Singh:2024vwb} 
While the leading order calculations suggest a strong enhancement at large angles, recent efforts towards a higher order description reveal an $\cal O$(1) deviation suppressing the contribution in this regime~\cite{Barata:2023bhh}. Recent theoretical efforts towards a systematically improvable effective field theory framework further indicate the importance of resummation in the small angle region for the two-point correlator~\cite{Singh:2024vwb}.

It was additionally proposed to study energy correlator observables using $\gamma$-tagged jets to not only mitigate the leading jet bias effects, but also to provide a unique approach to exposing a possible medium response, or the 'wake' generated in the medium by the passage of the jet. Utilizing the Hybrid model, the authors~\cite{Bossi:2024qho} studied the shape dependence of the full 3-point energy correlator. By looking at the largest separation between the three directions of the triangle, it was shown that the wake could be exposed in the ratio at large values of the observable. A very first analysis of the two-point energy correlator in the medium was recently presented by the CMS collaboration, revealing interesting modifications at large angular scales, see~\cite{Mainz_Viinikainen} and CMS PAS HIN-23-004 for details. Future theoretical and phenomenological efforts are needed to completely disentangle the different medium effects specifically at large angles.  

Since the leading jet bias effect reduces the sensitivity of inclusive measurements to jet quenching, multivariate approaches are being considered to provide increased sensitivity. 
In a recent work in Ref~\cite{Cunqueiro:2023vxl}, the authors showed that by selecting the hardest splitting in a jet above a certain transverse momentum scale $k_t^{\rm min}$ and studying its angular distribution, different energy loss mechanisms in the medium may be isolated.  
Sufficiently large values of $k_t^{\rm min}$ select the regime dominated by vacuum-like splittings and energy loss with little-to-no modification of the internal structure of jets. Varying $k_t^{\rm min}$ to lower values enhances the sensitivity of the observable towards other medium effects such as color coherence
while even smaller $k_t^{\rm min} \sim {\cal O}(1\, {\rm GeV})$ receive contribution from various competing effects such as multiple soft scatterings, medium recoil as well as non-perturbative hadronization effects which are hard to disentangle. 

In another recent work of Ref.~\cite{Mehtar-Tani:2024jtd}, the authors showed that by making use of inclusive jet $R_{AA}$ measurements along with jet azimuthal anisotropies over various centrality bins could provide a distinct probe of the coherence physics in the medium. Here the azimuthal anisotropy is caused by the fact that particles oriented in the direction in which medium is shorter suffer less energy loss compared to those in the direction where the medium is longer. A key idea of this framework is the determination of the resolved phase space in a jet where all resolved partons/subjets undergo quenching independently. The amount of resolved phase-space is determined by the physics of color decoherence in the medium with dipoles with an angle smaller than the critical angle $\theta_c$ loosing energy as a single color source. As the critical angle depends on the distance traversed by the jet in the medium, a study of jet quenching observables as a function of centrality may provide a handle into the coherence physics in the medium.

Finally, a promising novel direction is to use asymmetry observables that are affected by the finite flow velocity of the medium and study its effect on jet observables. It has been shown that the parton energy loss in the medium is not only impacted by the energy density of the medium but will also depend on the strength and direction of the collective flow field~\cite{Armesto:2004pt}. Recent developments \cite{Sadofyev:2021ohn,Barata:2022krd,Andres:2022ndd,Barata:2023qds,Kuzmin:2023hko} have established a first principles formulation of quenching in the presence of density gradients and flow.
Particles oriented in the direction of medium flow will experience a preferential $p_T$ broadening known as {\it drift}~\cite{Antiporda:2021hpk, ECT_bahder}. Efforts have also been made to study the impact of spatial inhomogeneities in the medium~\cite{PhysRevLett.125.122301, ECT_Sievert}. While most current formulations studied the effects of flow on hard partons, its impact on jet substructure measurements was addressed in \cite{Barata:2023zqg}. 

The presence of multiple scales in the evolution of jets make them an interesting probe of the medium dynamics but also complicates any attempts towards extraction of the medium scales of interest. Several approaches were proposed to study the dynamics of medium-induced modifications of parton showers. Generically speaking, it was identified that observables with: (a) $\gamma$/Z-tagged jet measurements, (b) well controlled vacuum baseline and (c)  reduced sensitivity to underlying event as well as medium response; are of potential interest.  

\section{Background considerations}
\label{sec:background}
 
Heavy ion collisions are a dense environment where a large number of particles are produced, most of which are not associated to a hard scattering event. In presently available data analyses, the $p_\mathrm{T}$-density of the background is estimated on an event-by-event basis using $\eta-\phi$ areas outside the jet cone and  subtracted, using either an area-based \cite{Soyez:2012hv,Cacciari:2010te} or the constituent-based method \cite{Berta:2014eza,Berta:2019hnj}. The remaining statistical fluctuations of the background level inside the jet cone lead to a significant smearing of the measured jet momentum and are generally corrected for using unfolding methods. More recently, machine-learning techniques have been used to develop a subtraction method for the jet momentum which reduces fluctuations below the statistical limit by incorporating multiplicity information in the calculation of the correction \cite{Haake:2018hqn,ALICE:2023waz,Mengel:2024fcl}.

While the effect of background and background fluctuations on most inclusive jet-shape variables is well-understood, observables that introduce additional cuts or selections may have additional sensitivity to background fluctuations. For example, upward fluctuations of the local background may promote a subjet above the grooming threshold in softdrop-based measurements, leading to an increase of apparent splitting rate in particular at large $R$, where the relative contribution of the background to the total energy flow is largest \cite{Milhano:2017nzm,Andrews:2018jcm}. 

The relative contributions from underlying event background are largest at low momenta and/or large angles to the jet axis, which makes measurements in those regimes the most sensitive to background fluctuations. Genuine medium effects, produced by the transfer of momentum to medium constituents (recoil, or wake effects) are expected to be strongest at momenta close to the thermal scale, where background fluctuations are large. The study of medium response therefore requires excellent control of the background fluctuations. Or, in other words: special care is needed to disentangle background fluctuations from medium response.
A detailed study, presented at the workshop \cite{ECT_Goncalves}, showed that the ability to distinguish between modified jets and vacuum-like ones with the Machine Learning approach of \cite{CrispimRomao:2023ssj} is enhanced by the presence of medium response and that this enhancement survives when background fluctuations are accounted for.

\section*{Acknowledgements}

We thank Alba Soto-Ontoso, Hannah Bossi, Jack Holguin,  
Konrad Tywoniuk, Leticia Cunqueiro, Matthew Sievert and Yen-Jie Lee for helpful discussions during the workshop. 
This work is a result of the activities of the Networking Activity \textit{'NA3-Jet-QGP: Quark-Gluon Plasma characterisation with jets'} of STRONG-2020 "The strong interaction at the frontier of knowledge: fundamental research and applications" which has received funding from the European Union’s Horizon 2020 research and innovation programme under grant agreement No 824093.

\printbibliography

@article{Mehtar-Tani:2019rrk,
    author = "Mehtar-Tani, Yacine and Soto-Ontoso, Alba and Tywoniuk, Konrad",
    title = "{Dynamical grooming of QCD jets}",
    eprint = "1911.00375",
    archivePrefix = "arXiv",
    primaryClass = "hep-ph",
    doi = "10.1103/PhysRevD.101.034004",
    journal = "Phys. Rev. D",
    volume = "101",
    number = "3",
    pages = "034004",
    year = "2020"
}

@article{Mehtar-Tani:2024jtd,
    author = "Mehtar-Tani, Yacine and Pablos, Daniel and Tywoniuk, Konrad",
    title = "{Jet suppression and azimuthal anisotropy from RHIC to LHC}",
    eprint = "2402.07869",
    archivePrefix = "arXiv",
    primaryClass = "hep-ph",
    doi = "10.1103/PhysRevD.110.014009",
    journal = "Phys. Rev. D",
    volume = "110",
    number = "1",
    pages = "014009",
    year = "2024"
}

@article{Haake:2018hqn,
    author = {Haake, R\"udiger and Loizides, Constantin},
    title = "{Machine Learning based jet momentum reconstruction in heavy-ion collisions}",
    eprint = "1810.06324",
    archivePrefix = "arXiv",
    primaryClass = "nucl-ex",
    doi = "10.1103/PhysRevC.99.064904",
    journal = "Phys. Rev. C",
    volume = "99",
    number = "6",
    pages = "064904",
    year = "2019"
}

@article{ALICE:2023waz,
    author = "Acharya, Shreyasi and others",
    collaboration = "ALICE",
    title = "{Measurement of the radius dependence of charged-particle jet suppression in Pb\textendash{}Pb collisions at sNN=5.02TeV}",
    eprint = "2303.00592",
    archivePrefix = "arXiv",
    primaryClass = "nucl-ex",
    reportNumber = "CERN-EP-2023-027",
    doi = "10.1016/j.physletb.2023.138412",
    journal = "Phys. Lett. B",
    volume = "849",
    pages = "138412",
    year = "2024"
}

@article{Mengel:2024fcl,
    author = "Mengel, Tanner and Steffanic, Patrick and Hughes, Charles and Da Silva, Antonio Carlos Oliveira and Nattrass, Christine",
    title = "{Multiplicity Based Background Subtraction for Jets in Heavy Ion Collisions}",
    eprint = "2402.10945",
    archivePrefix = "arXiv",
    primaryClass = "hep-ex",
    month = "2",
    year = "2024"
}

@article{Cunqueiro:2023vxl,
    author = "Cunqueiro, Leticia and Pablos, Daniel and Soto-Ontoso, Alba and Spousta, Martin and Takacs, Adam and Verweij, Marta",
    title = "{Isolating perturbative QCD splittings in heavy-ion collisions}",
    eprint = "2311.07643",
    archivePrefix = "arXiv",
    primaryClass = "hep-ph",
    reportNumber = "CERN-TH-2023-212",
    doi = "10.1103/PhysRevD.110.014015",
    journal = "Phys. Rev. D",
    volume = "110",
    number = "1",
    pages = "014015",
    year = "2024"
}

@article{Pablos:2022mrx,
    author = "Pablos, Daniel and Soto-Ontoso, Alba",
    title = "{Pushing forward jet substructure measurements in heavy-ion collisions}",
    eprint = "2210.07901",
    archivePrefix = "arXiv",
    primaryClass = "hep-ph",
    reportNumber = "CERN-TH-2022-164",
    doi = "10.1103/PhysRevD.107.094003",
    journal = "Phys. Rev. D",
    volume = "107",
    number = "9",
    pages = "094003",
    year = "2023"
}

@article{Apolinario:2020uvt,
    author = "Apolin\'ario, Liliana and Cordeiro, Andr\'e and Zapp, Korinna",
    title = "{Time reclustering for jet quenching studies}",
    eprint = "2012.02199",
    archivePrefix = "arXiv",
    primaryClass = "hep-ph",
    reportNumber = "LU TP 20-52",
    doi = "10.1140/epjc/s10052-021-09346-8",
    journal = "Eur. Phys. J. C",
    volume = "81",
    number = "6",
    pages = "561",
    year = "2021"
}

@article{CrispimRomao:2023ssj,
    author = "Crispim Rom\~ao, Miguel and Milhano, Jos\'e Guilherme and van Leeuwen, Marco",
    title = "{Jet substructure observables for jet quenching in quark gluon plasma: A machine learning driven analysis}",
    eprint = "2304.07196",
    archivePrefix = "arXiv",
    primaryClass = "hep-ph",
    doi = "10.21468/SciPostPhys.16.1.015",
    journal = "SciPost Phys.",
    volume = "16",
    number = "1",
    pages = "015",
    year = "2024"
}

@article{Salgado:2003gb,
    author = "Salgado, Carlos A. and Wiedemann, Urs Achim",
    title = "{Calculating quenching weights}",
    eprint = "hep-ph/0302184",
    archivePrefix = "arXiv",
    reportNumber = "CERN-TH-2003-030",
    doi = "10.1103/PhysRevD.68.014008",
    journal = "Phys. Rev. D",
    volume = "68",
    pages = "014008",
    year = "2003"
}

@article{Armesto:2011ht,
    author = "Armesto, Nestor and others",
    title = "{Comparison of Jet Quenching Formalisms for a Quark-Gluon Plasma 'Brick'}",
    eprint = "1106.1106",
    archivePrefix = "arXiv",
    primaryClass = "hep-ph",
    doi = "10.1103/PhysRevC.86.064904",
    journal = "Phys. Rev. C",
    volume = "86",
    pages = "064904",
    year = "2012"
}

@article{ATLAS:2018dgb,
    author = "Aaboud, Morad and others",
    collaboration = "ATLAS",
    title = "{Measurement of photon\textendash{}jet transverse momentum correlations in 5.02 TeV Pb + Pb and $pp$ collisions with ATLAS}",
    eprint = "1809.07280",
    archivePrefix = "arXiv",
    primaryClass = "nucl-ex",
    reportNumber = "CERN-EP-2018-196",
    doi = "10.1016/j.physletb.2018.12.023",
    journal = "Phys. Lett. B",
    volume = "789",
    pages = "167--190",
    year = "2019"
}

@article{CMS:2018mqn,
    author = "Sirunyan, Albert M and others",
    collaboration = "CMS",
    title = "{Observation of Medium-Induced Modifications of Jet Fragmentation in Pb-Pb Collisions at $\sqrt{s_{NN}}=$ 5.02  TeV Using Isolated Photon-Tagged Jets}",
    eprint = "1801.04895",
    archivePrefix = "arXiv",
    primaryClass = "hep-ex",
    reportNumber = "CMS-HIN-16-014, CERN-EP-2017-337",
    doi = "10.1103/PhysRevLett.121.242301",
    journal = "Phys. Rev. Lett.",
    volume = "121",
    number = "24",
    pages = "242301",
    year = "2018"
}

@article{Bossi:2024qho,
    author = "Bossi, Hannah and Kudinoor, Arjun Srinivasan and Moult, Ian and Pablos, Daniel and Rai, Ananya and Rajagopal, Krishna",
    title = "{Imaging the Wakes of Jets with Energy-Energy-Energy Correlators}",
    eprint = "2407.13818",
    archivePrefix = "arXiv",
    primaryClass = "hep-ph",
    reportNumber = "MIT-CTP-5739",
    month = "7",
    year = "2024"
}

@article{CMS:2018jco,
    author = "Sirunyan, Albert M and others",
    collaboration = "CMS",
    title = "{Jet Shapes of Isolated Photon-Tagged Jets in Pb-Pb and pp Collisions at $\sqrt{s_\mathrm{NN}} =$ 5.02  TeV}",
    eprint = "1809.08602",
    archivePrefix = "arXiv",
    primaryClass = "hep-ex",
    reportNumber = "CMS-HIN-18-006, CERN-EP-2018-249",
    doi = "10.1103/PhysRevLett.122.152001",
    journal = "Phys. Rev. Lett.",
    volume = "122",
    number = "15",
    pages = "152001",
    year = "2019"
}

@article{Singh:2024vwb,
    author = "Singh, Balbeer and Vaidya, Varun",
    title = "{Factorization for energy-energy correlator in heavy ion collision}",
    eprint = "2408.02753",
    archivePrefix = "arXiv",
    primaryClass = "hep-ph",
    month = "8",
    year = "2024"
}

@article{Zhang:2023oid,
    author = "Zhang, Shan-Liang and Wang, Enke and Xing, Hongxi and Zhang, Ben-Wei",
    title = "{Flavor dependence of jet quenching in heavy-ion collisions from a Bayesian analysis}",
    eprint = "2303.14881",
    archivePrefix = "arXiv",
    primaryClass = "hep-ph",
    doi = "10.1016/j.physletb.2024.138549",
    journal = "Phys. Lett. B",
    volume = "850",
    pages = "138549",
    year = "2024"
}

@article{Vaidya:2020lih,
    author = "Vaidya, Varun",
    title = "{Effective Field Theory for jet substructure in heavy ion collisions}",
    eprint = "2010.00028",
    archivePrefix = "arXiv",
    primaryClass = "hep-ph",
    reportNumber = "MIT-CTP 5243",
    doi = "10.1007/JHEP11(2021)064",
    journal = "JHEP",
    volume = "11",
    pages = "064",
    year = "2021"
}

@article{Vaidya:2020cyi,
    author = "Vaidya, Varun and Yao, Xiaojun",
    title = "{Transverse momentum broadening of a jet in quark-gluon plasma: an open quantum system EFT}",
    eprint = "2004.11403",
    archivePrefix = "arXiv",
    primaryClass = "hep-ph",
    reportNumber = "MIT-CTP/5187",
    doi = "10.1007/JHEP10(2020)024",
    journal = "JHEP",
    volume = "10",
    pages = "024",
    year = "2020"
}

@article{Cao:2017hhk,
    author = "Cao, Shanshan and Luo, Tan and Qin, Guang-You and Wang, Xin-Nian",
    title = "{Heavy and light flavor jet quenching at RHIC and LHC energies}",
    eprint = "1703.00822",
    archivePrefix = "arXiv",
    primaryClass = "nucl-th",
    doi = "10.1016/j.physletb.2017.12.023",
    journal = "Phys. Lett. B",
    volume = "777",
    pages = "255--259",
    year = "2018"
}

@article{Barata:2023zqg,
    author = "Barata, Jo\~ao and Milhano, Jos\'e Guilherme and Sadofyev, Andrey V.",
    title = "{Picturing QCD jets in anisotropic matter: from jet shapes to energy energy correlators}",
    eprint = "2308.01294",
    archivePrefix = "arXiv",
    primaryClass = "hep-ph",
    doi = "10.1140/epjc/s10052-024-12514-1",
    journal = "Eur. Phys. J. C",
    volume = "84",
    number = "2",
    pages = "174",
    year = "2024"
}

@article{Armesto:2004pt,
    author = "Armesto, Nestor and Salgado, Carlos A. and Wiedemann, Urs Achim",
    title = "{Measuring the collective flow with jets}",
    eprint = "hep-ph/0405301",
    archivePrefix = "arXiv",
    reportNumber = "CERN-PH-TH-2004-96",
    doi = "10.1103/PhysRevLett.93.242301",
    journal = "Phys. Rev. Lett.",
    volume = "93",
    pages = "242301",
    year = "2004"
}

@article{Barata:2022krd,
    author = "Barata, Jo\~ao and Sadofyev, Andrey V. and Salgado, Carlos A.",
    title = "{Jet broadening in dense inhomogeneous matter}",
    eprint = "2202.08847",
    archivePrefix = "arXiv",
    primaryClass = "hep-ph",
    doi = "10.1103/PhysRevD.105.114010",
    journal = "Phys. Rev. D",
    volume = "105",
    number = "11",
    pages = "114010",
    year = "2022"
}

@article{Antiporda:2021hpk,
    author = "Antiporda, Logan and Bahder, Joseph and Rahman, Hasan and Sievert, Matthew D.",
    title = "{Jet drift and collective flow in heavy-ion collisions}",
    eprint = "2110.03590",
    archivePrefix = "arXiv",
    primaryClass = "hep-ph",
    doi = "10.1103/PhysRevD.105.054025",
    journal = "Phys. Rev. D",
    volume = "105",
    number = "5",
    pages = "054025",
    year = "2022"
}

@article{Milhano:2017nzm,
    author = "Milhano, Guilherme and Wiedemann, Urs Achim and Zapp, Korinna Christine",
    title = "{Sensitivity of jet substructure to jet-induced medium response}",
    eprint = "1707.04142",
    archivePrefix = "arXiv",
    primaryClass = "hep-ph",
    reportNumber = "CERN-TH-2017-150, MCNET-17-12",
    doi = "10.1016/j.physletb.2018.01.029",
    journal = "Phys. Lett. B",
    volume = "779",
    pages = "409--413",
    year = "2018"
}

@article{Andrews:2018jcm,
    author = "Andrews, Harry Arthur and others",
    title = "{Novel tools and observables for jet physics in heavy-ion collisions}",
    eprint = "1808.03689",
    archivePrefix = "arXiv",
    primaryClass = "hep-ph",
    reportNumber = "CERN-TH-2018-186, LU-TP 18-14, IFJPAN-IV-2018-8, MCNET-18-19",
    doi = "10.1088/1361-6471/ab7cbc",
    journal = "J. Phys. G",
    volume = "47",
    number = "6",
    pages = "065102",
    year = "2020"
}

@article{Chen:2020adz,
    author = "Chen, Hao and Moult, Ian and Zhu, Hua Xing",
    title = "{Quantum Interference in Jet Substructure from Spinning Gluons}",
    eprint = "2011.02492",
    archivePrefix = "arXiv",
    primaryClass = "hep-ph",
    doi = "10.1103/PhysRevLett.126.112003",
    journal = "Phys. Rev. Lett.",
    volume = "126",
    number = "11",
    pages = "112003",
    year = "2021"
}

@article{Gao:2023ivm,
    author = "Gao, Anjie and Li, Hai Tao and Moult, Ian and Zhu, Hua Xing",
    title = "{The Transverse Energy-Energy Correlator at Next-to-Next-to-Next-to-Leading Logarithm}",
    eprint = "2312.16408",
    archivePrefix = "arXiv",
    primaryClass = "hep-ph",
    reportNumber = "MIT-CTP 5662",
    month = "12",
    year = "2023"
}

@article{Dixon:2019uzg,
    author = "Dixon, Lance J. and Moult, Ian and Zhu, Hua Xing",
    title = "{Collinear limit of the energy-energy correlator}",
    eprint = "1905.01310",
    archivePrefix = "arXiv",
    primaryClass = "hep-ph",
    reportNumber = "SLAC-PUB-17427, SLAC--PUB--17427",
    doi = "10.1103/PhysRevD.100.014009",
    journal = "Phys. Rev. D",
    volume = "100",
    number = "1",
    pages = "014009",
    year = "2019"
}

@article{PhysRevLett.125.122301,
  title = {Gradient Tomography of Jet Quenching in Heavy-Ion Collisions},
  author = {He, Yayun and Pang, Long-Gang and Wang, Xin-Nian},
  journal = {Phys. Rev. Lett.},
  volume = {125},
  issue = {12},
  pages = {122301},
  numpages = {5},
  year = {2020},
  month = {Sep},
  publisher = {American Physical Society},
  doi = {10.1103/PhysRevLett.125.122301},
  url = {https://link.aps.org/doi/10.1103/PhysRevLett.125.122301}
}

@conference{ECT_bahder,
    author = {Bahder, Joseph},
    title = "{Anisotropic jet broadening and the $R_{AA} X v_2$ puzzle }",
    booktitle = "ECT* Jet quenching tools workshop",
    year = 2024,
    url =  {https://indico.ectstar.eu/event/198/contributions/4382/attachments/2876/4021/JPB_Jet_Drift_ECTS_2024.02.08.pdf}
}

@conference{ECT_cunqueiro,
    author = {Cunqueiro Mendez, Leticia},
    title = "{What to measure}",
    booktitle = "ECT* Jet quenching tools workshop",
    year = 2024,
    url =  {https://indico.ectstar.eu/event/198/contributions/4458/attachments/2859/3997/WTM_Cunqueiro.pdf}
}

@article{Andres:2022,
    author = "Andres, Carlota and Dominguez, Fabio and Kunnawalkam Elayavalli, Raghav and Holguin, Jack and Marquet, Cyrille and Moult, Ian",
    title = "{Resolving the Scales of the Quark-Gluon Plasma with Energy Correlators}",
    eprint = "2209.11236",
    archivePrefix = "arXiv",
    primaryClass = "hep-ph",
    doi = "10.1103/PhysRevLett.130.262301",
    journal = "Phys. Rev. Lett.",
    volume = "130",
    number = "26",
    pages = "262301",
    year = "2023"
}

@conference{ECT_Goncalves,
    author = {Gon\c calves, Jo\~ao},
    title = "{Apples to Apples in Jet Quenching}",
 booktitle = "ECT* Jet quenching tools workshop",
    year = 2024,
    url =  {https://indico.ectstar.eu/event/198/contributions/4404/attachments/2895/4055/A_Goncalves_Trento.pdf}
}

@article{Andres:2023xwr,
    author = "Andres, Carlota and Dominguez, Fabio and Holguin, Jack and Marquet, Cyrille and Moult, Ian",
    title = "{A coherent view of the quark-gluon plasma from energy correlators}",
    eprint = "2303.03413",
    archivePrefix = "arXiv",
    primaryClass = "hep-ph",
    doi = "10.1007/JHEP09(2023)088",
    journal = "JHEP",
    volume = "09",
    pages = "088",
    year = "2023"
}

@misc{Andres:2024ksi,
    author = "Andres, Carlota and Dominguez, Fabio and Holguin, Jack and Marquet, Cyrille and Moult, Ian",
    title = "{Towards an Interpretation of the First Measurements of Energy Correlators in the Quark-Gluon Plasma}",
    eprint = "2407.07936",
    archivePrefix = "arXiv",
    primaryClass = "hep-ph",
    month = "7",
    year = "2024"
}

@article{Casalderrey-Solana:2019ubu,
    author = "Casalderrey-Solana, J. and Milhano, G. and Pablos, D. and Rajagopal, K.",
    title = "{Modification of Jet Substructure in Heavy Ion Collisions as a Probe of the Resolution Length of Quark-Gluon Plasma}",
    eprint = "1907.11248",
    archivePrefix = "arXiv",
    primaryClass = "hep-ph",
    doi = "10.1007/JHEP01(2020)044",
    journal = "JHEP",
    volume = "01",
    pages = "044",
    year = "2020"
}

@article{Barata:2023bhh,
    author = "Barata, Jo\~ao and Caucal, Paul and Soto-Ontoso, Alba and Szafron, Robert",
    title = "{Advancing the understanding of energy-energy correlators in heavy-ion collisions}",
    eprint = "2312.12527",
    archivePrefix = "arXiv",
    primaryClass = "hep-ph",
    month = "12",
    year = "2023"
}

@conference{ECT_Sievert,
    author = {Sievert D., Matthew},
    title = "{Survey of observables : medium-induced asymmetries}",
    booktitle = "ECT* Jet quenching tools workshop",
    year = 2024,
    url =  {https://indico.ectstar.eu/event/198/contributions/4460/attachments/2850/3984/Sievert_Overview.pdf}
}

@conference{Mainz_Viinikainen,
    author = {Viinikainen, Jussi},
    title = "{Energy-energy correlators from PbPb and pp collisions at 5.02 TeV with CMS}",
    booktitle = "{Energy correlators at the collider frontier}", 
    year = 2024,
    url =  {https://indico.mitp.uni-mainz.de/event/358/contributions/4984/attachments/3595/4660/jussi_energyEnergyCorrelators_mainz2024.pdf}
}

@conference{ECT_lee,
    author = "Lee, Yen-Jie",
    title = "{New jet quenching tools to explore equilibrium and non-equilibrium dynamics in heavy-ion collisions}",
    booktitle = "ECT* Jet quenching tools workshop",
    year = 2024,
    url = {https://indico.ectstar.eu/event/198/contributions/4459/attachments/2854/3992/20240112-yenjie-Obs-Trento-v2.pdf}
}

@article{Soyez:2012hv,
    author = "Soyez, Gregory and Salam, Gavin P. and Kim, Jihun and Dutta, Souvik and Cacciari, Matteo",
    title = "{Pileup subtraction for jet shapes}",
    eprint = "1211.2811",
    archivePrefix = "arXiv",
    primaryClass = "hep-ph",
    reportNumber = "CERN-PH-TH-2012-300",
    doi = "10.1103/PhysRevLett.110.162001",
    journal = "Phys. Rev. Lett.",
    volume = "110",
    number = "16",
    pages = "162001",
    year = "2013"
}

@article{Cacciari:2010te,
    author = "Cacciari, Matteo and Rojo, Juan and Salam, Gavin P. and Soyez, Gregory",
    title = "{Jet Reconstruction in Heavy Ion Collisions}",
    eprint = "1010.1759",
    archivePrefix = "arXiv",
    primaryClass = "hep-ph",
    reportNumber = "CERN-PH-TH-2010-223",
    doi = "10.1140/epjc/s10052-011-1539-z",
    journal = "Eur. Phys. J. C",
    volume = "71",
    pages = "1539",
    year = "2011"
}

@article{Berta:2014eza,
    author = "Berta, Peter and Spousta, Martin and Miller, David W. and Leitner, Rupert",
    title = "{Particle-level pileup subtraction for jets and jet shapes}",
    eprint = "1403.3108",
    archivePrefix = "arXiv",
    primaryClass = "hep-ex",
    doi = "10.1007/JHEP06(2014)092",
    journal = "JHEP",
    volume = "06",
    pages = "092",
    year = "2014"
}

@article{Cunqueiro:2021wls,
    author = "Cunqueiro, Leticia and Sickles, Anne M.",
    title = "{Studying the QGP with Jets at the LHC and RHIC}",
    eprint = "2110.14490",
    archivePrefix = "arXiv",
    primaryClass = "nucl-ex",
    doi = "10.1016/j.ppnp.2022.103940",
    journal = "Prog. Part. Nucl. Phys.",
    volume = "124",
    pages = "103940",
    year = "2022"
}

@article{Chien:2015hda,
    author = "Chien, Yang-Ting and Vitev, Ivan",
    title = "{Towards the understanding of jet shapes and cross sections in heavy ion collisions using soft-collinear effective theory}",
    eprint = "1509.07257",
    archivePrefix = "arXiv",
    primaryClass = "hep-ph",
    doi = "10.1007/JHEP05(2016)023",
    journal = "JHEP",
    volume = "05",
    pages = "023",
    year = "2016"
}

@article{Yang:2023dwc,
    author = "Yang, Zhong and He, Yayun and Moult, Ian and Wang, Xin-Nian",
    title = "{Probing the Short-Distance Structure of the Quark-Gluon Plasma with Energy Correlators}",
    eprint = "2310.01500",
    archivePrefix = "arXiv",
    primaryClass = "hep-ph",
    doi = "10.1103/PhysRevLett.132.011901",
    journal = "Phys. Rev. Lett.",
    volume = "132",
    number = "1",
    pages = "011901",
    year = "2024"
}

@article{Salgado:2003rv,
    author = "Salgado, Carlos A. and Wiedemann, Urs Achim",
    title = "{Medium modification of jet shapes and jet multiplicities}",
    eprint = "hep-ph/0310079",
    archivePrefix = "arXiv",
    reportNumber = "CERN-TH-2003-244",
    doi = "10.1103/PhysRevLett.93.042301",
    journal = "Phys. Rev. Lett.",
    volume = "93",
    pages = "042301",
    year = "2004"
}

@article{Caucal:2019uvr,
    author = "Caucal, P. and Iancu, E. and Soyez, G.",
    title = "{Deciphering the $z_g$ distribution in ultrarelativistic heavy ion collisions}",
    eprint = "1907.04866",
    archivePrefix = "arXiv",
    primaryClass = "hep-ph",
    doi = "10.1007/JHEP10(2019)273",
    journal = "JHEP",
    volume = "10",
    pages = "273",
    year = "2019"
}

@article{Casalderrey-Solana:2014bpa,
    author = "Casalderrey-Solana, Jorge and Gulhan, Doga Can and Milhano, Jos\'e Guilherme and Pablos, Daniel and Rajagopal, Krishna",
    title = "{A Hybrid Strong/Weak Coupling Approach to Jet Quenching}",
    eprint = "1405.3864",
    archivePrefix = "arXiv",
    primaryClass = "hep-ph",
    reportNumber = "MIT-CTP-4550, CERN-PH-TH-2014-089, ICCUB-14-051",
    doi = "10.1007/JHEP09(2015)175",
    journal = "JHEP",
    volume = "10",
    pages = "019",
    year = "2014",
    note = "[Erratum: JHEP 09, 175 (2015)]"
}

@article{Zapp:2008gi,
    author = "Zapp, Korinna and Ingelman, Gunnar and Rathsman, Johan and Stachel, Johanna and Wiedemann, Urs Achim",
    title = "{A Monte Carlo Model for 'Jet Quenching'}",
    eprint = "0804.3568",
    archivePrefix = "arXiv",
    primaryClass = "hep-ph",
    reportNumber = "CERN-PH-TH-2008-067",
    doi = "10.1140/epjc/s10052-009-0941-2",
    journal = "Eur. Phys. J. C",
    volume = "60",
    pages = "617--632",
    year = "2009"
}

@article{Mehtar-Tani:2021fud,
    author = "Mehtar-Tani, Yacine and Pablos, Daniel and Tywoniuk, Konrad",
    title = "{Cone-Size Dependence of Jet Suppression in Heavy-Ion Collisions}",
    eprint = "2101.01742",
    archivePrefix = "arXiv",
    primaryClass = "hep-ph",
    doi = "10.1103/PhysRevLett.127.252301",
    journal = "Phys. Rev. Lett.",
    volume = "127",
    number = "25",
    pages = "252301",
    year = "2021"
}

@article{Caucal:2021cfb,
    author = "Caucal, Paul and Soto-Ontoso, Alba and Takacs, Adam",
    title = "{Dynamically groomed jet radius in heavy-ion collisions}",
    eprint = "2111.14768",
    archivePrefix = "arXiv",
    primaryClass = "hep-ph",
    doi = "10.1103/PhysRevD.105.114046",
    journal = "Phys. Rev. D",
    volume = "105",
    number = "11",
    pages = "114046",
    year = "2022"
}

@article{Budhraja:2023rgo,
    author = "Budhraja, Ankita and Sharma, Rishi and Singh, Balbeer",
    title = "{Medium modifications to jet angularities using SCET with Glauber gluons}",
    eprint = "2305.10237",
    archivePrefix = "arXiv",
    primaryClass = "hep-ph",
    month = "5",
    year = "2023"
}

@article{Cao:2020wlm,
    author = "Cao, Shanshan and Wang, Xin-Nian",
    title = "{Jet quenching and medium response in high-energy heavy-ion collisions: a review}",
    eprint = "2002.04028",
    archivePrefix = "arXiv",
    primaryClass = "hep-ph",
    doi = "10.1088/1361-6633/abc22b",
    journal = "Rept. Prog. Phys.",
    volume = "84",
    number = "2",
    pages = "024301",
    year = "2021"
}

@article{ATLAS:2018gwx,
    author = "Aaboud, Morad and others",
    collaboration = "ATLAS",
    title = "{Measurement of the nuclear modification factor for inclusive jets in Pb+Pb collisions at $\sqrt{s_\mathrm{NN}}=5.02$ TeV with the ATLAS detector}",
    eprint = "1805.05635",
    archivePrefix = "arXiv",
    primaryClass = "nucl-ex",
    reportNumber = "CERN-EP-2018-105",
    doi = "10.1016/j.physletb.2018.10.076",
    journal = "Phys. Lett. B",
    volume = "790",
    pages = "108--128",
    year = "2019"
}

@article{ATLAS:2014ipv,
    author = "Aad, Georges and others",
    collaboration = "ATLAS",
    title = "{Measurements of the Nuclear Modification Factor for Jets in Pb+Pb Collisions at $\sqrt{s_{\mathrm{NN}}}=2.76$ TeV with the ATLAS Detector}",
    eprint = "1411.2357",
    archivePrefix = "arXiv",
    primaryClass = "hep-ex",
    reportNumber = "CERN-PH-EP-2014-172",
    doi = "10.1103/PhysRevLett.114.072302",
    journal = "Phys. Rev. Lett.",
    volume = "114",
    number = "7",
    pages = "072302",
    year = "2015"
}

@article{CMS:2016uxf,
    author = "Khachatryan, Vardan and others",
    collaboration = "CMS",
    title = "{Measurement of inclusive jet cross sections in $pp$ and PbPb collisions at $\sqrt{s_{NN}}=$ 2.76 TeV}",
    eprint = "1609.05383",
    archivePrefix = "arXiv",
    primaryClass = "nucl-ex",
    reportNumber = "CMS-HIN-13-005, CERN-EP-2016-217",
    doi = "10.1103/PhysRevC.96.015202",
    journal = "Phys. Rev. C",
    volume = "96",
    number = "1",
    pages = "015202",
    year = "2017"
}

@article{ALICE:2019qyj,
    author = "Acharya, Shreyasi and others",
    collaboration = "ALICE",
    title = "{Measurements of inclusive jet spectra in pp and central Pb-Pb collisions at $\sqrt{s_{\rm{NN}}}$ = 5.02 TeV}",
    eprint = "1909.09718",
    archivePrefix = "arXiv",
    primaryClass = "nucl-ex",
    reportNumber = "CERN-EP-2019-200",
    doi = "10.1103/PhysRevC.101.034911",
    journal = "Phys. Rev. C",
    volume = "101",
    number = "3",
    pages = "034911",
    year = "2020"
}

@article{Muller:2012zq,
    author = "Muller, Berndt and Schukraft, Jurgen and Wyslouch, Boleslaw",
    title = "{First Results from Pb+Pb collisions at the LHC}",
    eprint = "1202.3233",
    archivePrefix = "arXiv",
    primaryClass = "hep-ex",
    reportNumber = "CERN-OPEN-2012-005",
    doi = "10.1146/annurev-nucl-102711-094910",
    journal = "Ann. Rev. Nucl. Part. Sci.",
    volume = "62",
    pages = "361--386",
    year = "2012"
}

@article{Muller:2006ee,
    author = "Muller, Berndt and Nagle, James L.",
    title = "{Results from the relativistic heavy ion collider}",
    eprint = "nucl-th/0602029",
    archivePrefix = "arXiv",
    doi = "10.1146/annurev.nucl.56.080805.140556",
    journal = "Ann. Rev. Nucl. Part. Sci.",
    volume = "56",
    pages = "93--135",
    year = "2006"
}

@article{Gyulassy:2004zy,
    author = "Gyulassy, Miklos and McLerran, Larry",
    editor = "Rischke, D. and Levin, G.",
    title = "{New forms of QCD matter discovered at RHIC}",
    eprint = "nucl-th/0405013",
    archivePrefix = "arXiv",
    doi = "10.1016/j.nuclphysa.2004.10.034",
    journal = "Nucl. Phys. A",
    volume = "750",
    pages = "30--63",
    year = "2005"
}

@article{Berta:2019hnj,
    author = "Berta, P. and Masetti, L. and Miller, D. W. and Spousta, M.",
    title = "{Pileup and Underlying Event Mitigation with Iterative Constituent Subtraction}",
    eprint = "1905.03470",
    archivePrefix = "arXiv",
    primaryClass = "hep-ph",
    doi = "10.1007/JHEP08(2019)175",
    journal = "JHEP",
    volume = "08",
    pages = "175",
    year = "2019"
}

@article{Brewer:2018dfs,
    author = "Brewer, Jasmine and Milhano, Jos\'e Guilherme and Thaler, Jesse",
    title = "{Sorting out quenched jets}",
    eprint = "1812.05111",
    archivePrefix = "arXiv",
    primaryClass = "hep-ph",
    reportNumber = "MIT-CTP/5089",
    doi = "10.1103/PhysRevLett.122.222301",
    journal = "Phys. Rev. Lett.",
    volume = "122",
    number = "22",
    pages = "222301",
    year = "2019"
}

@article{Sadofyev:2021ohn,
    author = "Sadofyev, Andrey V. and Sievert, Matthew D. and Vitev, Ivan",
    title = "{Ab~initio coupling of jets to collective flow in the opacity expansion approach}",
    eprint = "2104.09513",
    archivePrefix = "arXiv",
    primaryClass = "hep-ph",
    reportNumber = "LA-UR-21-21420",
    doi = "10.1103/PhysRevD.104.094044",
    journal = "Phys. Rev. D",
    volume = "104",
    number = "9",
    pages = "094044",
    year = "2021"
}

@article{Andres:2022ndd,
    author = "Andres, Carlota and Dominguez, Fabio and Sadofyev, Andrey V. and Salgado, Carlos A.",
    title = "{Jet broadening in flowing matter: Resummation}",
    eprint = "2207.07141",
    archivePrefix = "arXiv",
    primaryClass = "hep-ph",
    doi = "10.1103/PhysRevD.106.074023",
    journal = "Phys. Rev. D",
    volume = "106",
    number = "7",
    pages = "074023",
    year = "2022"
}

@article{Barata:2023qds,
    author = "Barata, Jo\~ao and Mayo L\'opez, Xo\'an and Sadofyev, Andrey V. and Salgado, Carlos A.",
    title = "{Medium induced gluon spectrum in dense inhomogeneous matter}",
    eprint = "2304.03712",
    archivePrefix = "arXiv",
    primaryClass = "hep-ph",
    doi = "10.1103/PhysRevD.108.034018",
    journal = "Phys. Rev. D",
    volume = "108",
    number = "3",
    pages = "034018",
    year = "2023"
}

@article{Kuzmin:2023hko,
    author = "Kuzmin, Matvey V. and Mayo L\'opez, Xo\'an and Reiten, Jared and Sadofyev, Andrey V.",
    title = "{Jet quenching in anisotropic flowing matter}",
    eprint = "2309.00683",
    archivePrefix = "arXiv",
    primaryClass = "hep-ph",
    doi = "10.1103/PhysRevD.109.014036",
    journal = "Phys. Rev. D",
    volume = "109",
    number = "1",
    pages = "014036",
    year = "2024"
}

@article{Cao:2022odi,
    author = "Cao, Shanshan and Qin, Guang-You",
    title = "{Medium Response and Jet\textendash{}Hadron Correlations in Relativistic Heavy-Ion Collisions}",
    eprint = "2211.16821",
    archivePrefix = "arXiv",
    primaryClass = "nucl-th",
    doi = "10.1146/annurev-nucl-112822-031317",
    journal = "Ann. Rev. Nucl. Part. Sci.",
    volume = "73",
    pages = "205--229",
    year = "2023"
}

@article{He:2015pra,
    author = "He, Yayun and Luo, Tan and Wang, Xin-Nian and Zhu, Yan",
    title = "{Linear Boltzmann Transport for Jet Propagation in the Quark-Gluon Plasma: Elastic Processes and Medium Recoil}",
    eprint = "1503.03313",
    archivePrefix = "arXiv",
    primaryClass = "nucl-th",
    doi = "10.1103/PhysRevC.91.054908",
    journal = "Phys. Rev. C",
    volume = "91",
    pages = "054908",
    year = "2015",
    note = "[Erratum: Phys.Rev.C 97, 019902 (2018)]"
}

\end{document}